\documentclass[reprint]{revtex4-1}
\usepackage{amsmath}
\usepackage{bm}
\usepackage{graphicx}

\graphicspath{{./FIGURES/}}

\begin{document}


\title{Mode space approach for tight-binding transport simulations in graphene nanoribbon field-effect transistors including phonon scattering} 



\author{R. Grassi}
\email[]{rgrassi@arces.unibo.it}
\author{A. Gnudi}
\author{I. Imperiale}
\author{E. Gnani}
\author{S. Reggiani}
\author{G. Baccarani}
\affiliation{ARCES - DEI, University of Bologna, Viale Risorgimento 2, 40136 Bologna, Italy}


\date{\today}

\begin{abstract}
In this paper, we present a mode space method for atomistic non-equilibrium Green's function simulations of armchair graphene nanoribbon FETs that includes electron-phonon scattering. With reference to both conventional and tunnel FET structures, we show that, in the ideal case of a smooth electrostatic potential, the modes can be decoupled in different groups without any loss of accuracy. Thus, inter-subband scattering due to electron-phonon interactions is properly accounted for, while the overall simulation time considerably improves with respect to real-space, with a speed-up factor of 40 for a $1.5$-nm-wide device. Such factor increases with the square of the device width. We also discuss the accuracy of two commonly used approximations of the scattering self-energies: the neglect of the off-diagonal entries in the mode-space expressions and the neglect of the Hermitian part of the retarded self-energy. While the latter is an acceptable approximation in most bias conditions, the former is somewhat inaccurate when the device is in the off-state and optical phonon scattering is essential in determining the current via band-to-band tunneling. Finally, we show that, in the presence of a disordered potential, a coupled mode space approach is necessary, but the results are still accurate compared to the real-space solution.
\end{abstract}


\maketitle 

\section{Introduction}
\label{introduction}
Graphene nanoribbons (GNRs) have been proposed in recent years as a possible replacement for silicon in the future generation of field effect transistors \cite{HanPRL2007,ChenPE2007}. Despite the outstanding challenges in producing GNRs with controlled width and edges and the resulting degradation of the graphene intrinsic mobility, devices with large on-off current ratio have been successfully demonstrated \cite{WangPRL2008}.

Theoretical performance of GNRFETs have been widely investigated by numerous simulation studies (e.g Refs.~\onlinecite{FioriEDL2007,OuyangTED2007,LiangJAP2007,YoonTED2008,ZhaoNL2009,GrassiJCE2009}). The state-of-the-art approach for modeling the electronic properties of GNRs is based on an atomistic tight-binding (TB) Hamiltonian with a $p_z$ orbital basis set. The transport problem is usually solved within the nonequilibrium Green's function formalism (NEGF) \cite{DattaQT2005}, which provides a rigorous framework for including incoherent scattering processes in the quantum description. TB simulations of GNRFETs including phonon scattering have been reported too \cite{OuyangAPL2008,YoonAPL2011,YoonAPL2012,AkhavanJAP2012}. Those simulations were performed using a real space (RS) approach. On the other hand, more efficient mode space (MS) methods would be preferable for use in intensive device simulations.

The MS approach is well established with reference to an effective mass (EM) Hamiltonian \cite{WangAPL2004,LuisierAPL2006}. It is based on the expansion of the Green's functions in terms of the transverse eigenfunctions (modes) and it is most efficient when quantum confinement is relatively strong in the transverse plane, so that only few of the lowest subbands are occupied. The so-called coupled mode space (CMS) is the general method, while a more efficient version, the uncoupled mode space (UMS), can be adopted when the transverse potential profile is slightly varying along the transport direction. The inclusion of electron-phonon scattering in the MS EM model is also well known \cite{JinAPL2006,PoliThesis2009,EspositoJCE2009,NikonovNH2009,AfzalianAPL2011}.

On the other hand, the MS approach is not generally applicable to a TB Hamiltonian. Due to the non-separable Hamiltonian, even when the potential is uniform along the longitudinal direction (so that the longitudinal wavevector $k$ is conserved), the wavefunctions at different $k$ in the same subband are in general different. In other words, by defining the modes as the transverse part of the wavefunctions at a particular $k$, the wavefunctions at the generic $k$ are a linear combination of them. This is similar to what occurs in the case of a $k \cdot p$ Hamiltonian \cite{ShinAPL2009}. Nevertheless, for the particular case of the TB Hamiltonian of GNRs with armchair edges, we have previously shown by numerical calculation that a MS method is actually possible, since each subband contains only few modes \cite{GrassiTNANO2011}. A formal derivation was given in Ref.~\onlinecite{ZhaoJAP2009} using analytically defined modes.

In this paper, we propose a novel MS method for the inclusion of graphene acoustic phonon (AP) and optical phonon (OP) scattering within the NEGF formalism, and test its accuracy with respect to RS in both the cases of smooth and disordered potentials. Two approximations commonly found in the literature for treating the scattering self-energy are also evaluated. The paper is organized as follows. Sec.~\ref{sec_mathematical} reviews the standard RS TB method as well as the MS TB method from Ref.~\onlinecite{GrassiTNANO2011}, modified so as to include phonon scattering. Simulation results of GNRFETs are presented in Sec.~\ref{sec_results} and conclusions are finally drawn in Sec.~\ref{sec_conclusions}.


\section{Mathematical model}
\label{sec_mathematical}
\subsection{Real space formulation}

\begin{figure}
\includegraphics[width=0.9\linewidth]{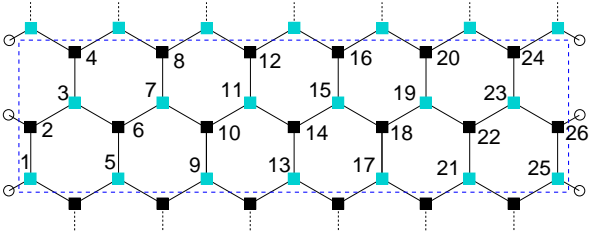}
\caption{\label{fig_ribbon}One-dimensional elementary cell of an $N_a = 13$ 
armchair GNR.} 
\end{figure}
The atomic structure of an armchair GNR unit cell is shown for reference in Fig.~\ref{fig_ribbon}. The GNR width is denoted by the number of dimer lines $N_a$. We adopt the TB Hamiltonian model proposed in Ref.~\onlinecite{SonPRL2006}, which is based on a set of orthogonal $p_z$ orbitals, one for each atom, with hopping integrals limited to first nearest-neighbor orbitals and of value $t$ for internal atom pairs, and $t(1+\delta)$ for atom pairs located along the edges of the GNR ($t = -2.7$~eV and $\delta=0.12$). The onsite energies are set equal to the value of the electrostatic potential energy at each atomic site. Unless stated otherwise, the electrostatic potential is calculated by self-consistently solving the 3D Poisson equation.

In order to calculate electron and hole densities within the NEGF formalism, the following equations need to be solved for the retarded, the lesser and the greater  Green's function matrices $\bm{G}^R$, $\bm{G}^<$ and $\bm{G}^>$, respectively, for each energy $E$
\begin{equation} 
\left[ (E+\text{i}0^+) \bm{I} - \bm{H}^d - \bm{\Sigma}^R(E) \right]  \bm{G}^R(E) = \bm{I}
\label{eqGr}
\end{equation}
\begin{equation}
\bm{G}^<(E) = \bm{G}^R(E)  \bm{\Sigma}^<(E) \bm{G}^A(E) 
\label{eqGlesser}
\end{equation}
\begin{equation}
\bm{G}^>(E) = \bm{G}^R(E) \bm{\Sigma}^>(E) \bm{G}^A(E) 
\label{eqGgreater}
\end{equation}
where $\bm{H}^d$ denotes the restriction of the Hamiltonian matrix $\bm{H}$ to the GNR portion inside
the simulation domain and $\bm{G}^A = \bm{G}^{R\dag}$. Besides, 
$\bm{\Sigma}^R(E) = \bm{\Sigma}^R_P(E) + \bm{\Sigma}^R_S(E) + \bm{\Sigma}^R_D(E)$ and similarly for $\bm{\Sigma}^>(E) $ and 
$\bm{\Sigma}^<(E)$, where 
$\bm{\Sigma}^{(...)}_{S/D}$ are the self-energies for the semi-infinite source and drain leads, which are assumed in thermodynamic equilibrium, 
while $\bm{\Sigma}^{(...)}_{P}$ account for phonon scattering. Calculation of both (\ref{eqGlesser}) and (\ref{eqGgreater}) is actually not needed, since either one of the two equations can be replaced by the identity $\bm{G}^>-\bm{G}^< = \bm{G}^R-\bm{G}^A$.

Phonon scattering is treated within the self-consistent Born approximation. The model accounts for both OP and AP scattering, the latter within the elastic and high temperature limit, resulting in the following expressions of the lesser and greater phonon self energies
\begin{eqnarray}
\bm{\Sigma}^<_P(E) = \left[ D_{ap} \bm{G}^<(E) \right. &+&  D_{op} (N_{op}+1) \bm{G}^<(E+\hbar \omega_{op}) + \nonumber\\
&+& \left. D_{op} N_{op} \bm{G}^<(E-\hbar \omega_{op}) \right] \circ \bm{I}
\label{eqSlesserPH}
\end{eqnarray}
\begin{eqnarray} 
\bm{\Sigma}^>_P(E) = \left[ D_{ap} \bm{G}^>(E) \right. &+&  D_{op} (N_{op}+1) \bm{G}^>(E - \hbar \omega_{op}) + \nonumber\\
&+& \left. D_{op} N_{op} \bm{G}^>(E + \hbar \omega_{op}) \right] \circ \bm{I}
\label{eqSgreaterPH}
\end{eqnarray}
where $\circ$ indicates element-by-element matrix multiplication, $N_{op}$ is the OP occupation number, and $D_{ap}$ and $D_{op}$ are given by
\begin{equation} 
D_{ap} = {{D_{ac}^2 k_B T} \over {4 m_c v_s^2}} \qquad , \qquad 
D_{op} = {{D_{o}^2 \hbar} \over {8 m_c \omega_{op}}} 
\label{eqDapDop}
\end{equation}
with $D_{ac} = 16\,$eV the AP deformation potential, $v_s = 2 \times 10^6\,$cm/s the sound velocity in graphene, $T$ the temperature, $m_c$ the carbon atomic mass, 
$D_{o} = 10^9\,$eV/cm and $\hbar \omega_{op} = 160\,$meV the zone-boundary OP deformation potential and energy, respectively. The retarded phonon self-energy is computed using the identity 
\begin{eqnarray} 
\bm{\Sigma}^R_P(E) = {{\text{i}}\over{2\pi}} \!\! \! \int\limits_{-\infty}^{+\infty}
{{\bm{\Sigma}^>_P(E^\prime) - \bm{\Sigma}^<_P(E^\prime)} 
\over{E+\text{i}0^+ - E^\prime}} \text{d}E^\prime =  \nonumber\\
\!\!\!\!\!\! = {{\bm{\Sigma}^>_P(E) - \bm{\Sigma}^<_P(E)} \over {2}} + 
\text{i} \text{P} \! \! \! \int\limits_{-\infty}^{+\infty} {{\bm{\Sigma}^>_P(E^\prime) - \bm{\Sigma}^<_P(E^\prime)} 
\over{2\pi(E - E^\prime})} \text{d}E^\prime 
\label{eqSrPH}
\end{eqnarray}
where the symbol $\text{P}$ stands for the principal part of the integral. Unless stated otherwise, we include the contribution to the principal part integral (Hermitian part of $\bm{\Sigma}^R_P$) due to AP scattering, while we neglect the one due to OP scattering. See Appendix~\ref{sec_hermitian_part} for details on the numerical implementation of the principal part integral.

The equations above, namely (\ref{eqGr})-(\ref{eqGgreater}) together with (\ref{eqSlesserPH}),
(\ref{eqSgreaterPH}) and (\ref{eqSrPH}), are iteratively solved. It is worth noticing that $\bm{H}^d$ and $\bm{\Sigma}^{(\ldots)}$ can be ordered in a block tridiagonal and block diagonal form, respectively, each diagonal block having the size of the number of atoms in a specific transverse layer, or slab, of the GNR. Hence, the Recursive Green Function algorithm (RGF) \cite{LakeJAP1997,SvizhenkoJAP2002} can be used to solve (\ref{eqGr})-(\ref{eqGgreater}). In this paper, we consider each slab to be made of four rows of atoms, i.e. equal to the GNR unit cell (Fig.~\ref{fig_ribbon}).

\subsection{Mode space formulation}
In this section the MS approach presented in Ref.~\onlinecite{GrassiTNANO2011}, and in that paper discussed with reference to coherent transport simulations of GNRs, is extended to include phonon scattering.

The MS formulation starts by defining a set of orthonormal vectors, or modes, $\bm{\phi}(i)$ for each slab $i$ of the device. We choose the modes of the generic slab as the eigenvectors computed at $k=0$ of a fictitious GNR obtained by the periodic repetition (potential energy included) of that slab along the longitudinal direction. In the absence of topological differences between the slabs (such as edge irregularities or internal vacancies), the modes are the solution of the eigenvalue problem
\begin{equation}
\left( \bm{H}_{i,i} + \bm{H}_{i,i+1} + \bm{H}_{i,i+1}^\dagger \right) \bm{\phi}^m(i) = \varepsilon^m \bm{\phi}^m(i)
\end{equation}
where $\bm{H}_{i,j}$ is the Hamiltonian block between slab $i$ and $j$.
Denoting by $\bm{v}(i)$ the matrix whose columns are the modes of layer $i$, i.e. $\bm{v}(i) = [ \bm{\phi}^1(i) \cdots \bm{\phi}^m(i) \cdots]$, a block-diagonal transformation matrix $\bm{V}$ satisfying $\bm{V}^\dagger \bm{V} = \bm{I}$ is constructed
\begin{equation}
\bm{V} = \left( \begin{array}{cccc}
\bm{v}(1) &         0 & \cdots &           0 \\
        0 & \bm{v}(2) & \ddots &      \vdots \\
   \vdots &    \ddots & \ddots &      \ddots \\
        0 &    \cdots & \ddots & \bm{v}(N_s)
\end{array} \right)
\label{eqV}
\end{equation}
with $N_s$ equal to the number of slabs.

In the following, we indicate with a tilde the MS matrices to distinguish them from the RS matrices. The CMS method consists in approximating $\bm{G}^R$ (and similarly for $\bm{G}^<$ and $\bm{G}^>$) as
\begin{equation}
\bm{G}^R(E) \simeq \bm{V} \bm{\widetilde{G}}^R(E) \bm{V}^\dagger
\label{eqGrMStoRS}
\end{equation}
where $\bm{\widetilde{G}}^R$ is the solution of
\begin{equation} 
\left[ (E+\text{i}0^+) \bm{I} - \bm{\widetilde{H}}^d - \bm{\widetilde{\Sigma}}^R(E) \right]  \bm{\widetilde{G}}^R(E) = \bm{I}
\label{eqGrMS}
\end{equation} 
with
\begin{eqnarray}
\bm{\widetilde{H}}^d &=& \bm{V}^\dagger \bm{H}^d \bm{V} \label{eqn_def_H_MS} \\
\bm{\widetilde{\Sigma}}^R(E) &=& \bm{V}^\dagger \bm{\Sigma}^R(E) \bm{V} 
\label{eqn_def_Sr_MS}
\end{eqnarray}
Eq.~(\ref{eqGrMS}) is the MS version of (\ref{eqGr}). Analogous considerations apply to the other equations. Eq.~(\ref{eqGrMStoRS}) would be exact if $\bm{V}$ was a square matrix. In practice, a mode truncation is performed so that $\bm{\widetilde{H}}^d$ and $\bm{\widetilde{\Sigma}}^R$ have a smaller size than their corresponding RS matrices and the solution of (\ref{eqGrMS}) instead of (\ref{eqGr}) is computationally advantageous. The computational time of the RGF algorithm scales as $O(N_y^3 N_s)$, where $N_y$ is the matrix block size \cite{SvizhenkoJAP2002}, which is equal to the slab size for RS ($2 N_a$ according to our choice of the slab size) and to the number $N_m$ of selected modes per slab for CMS. Thus, the speed-up of CMS compared to RS is a factor of the order of $(2 N_a/N_m)^3$.

An algorithm to select the modes to retain in the $\bm{V}$ matrix was presented in Ref.~\onlinecite{GrassiTNANO2011}. It is based on identifying, among the modes calculated with zero electrostatic potential, the minimum set of modes that allow to reproduce with sufficient accuracy the subbands that lie in the energy range of interest.
The mode indexes so identified are then used to select the actual modes calculated with the non-null electrostatic potential.
\begin{figure}
\includegraphics[width=\linewidth]{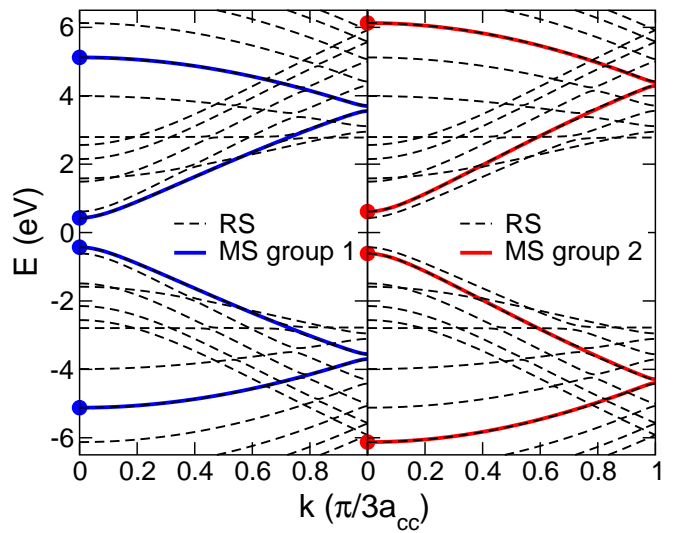}
\caption{\label{fig_bandems2}Subband structure of a $N_a=13$ GNR computed with RS (dashed lines) and MS (solid lines) using two different groups of modes (left and right). The modes included in each group are the ones that correspond to the eigenvalues at $k=0$ indicated with circles in each figure. The electrostatic potential is set to zero. $3 a_{cc}$ is the length of the GNR unit cell, with $a_{cc}$ the carbon interatomic distance.}
\end{figure}
As shown in Fig.~\ref{fig_bandems2} for a $N_a=13$ GNR, it turns out that each one of the lowest conduction subbands at zero potential can be well reproduced by just four modes, which correspond to the eigenvalues at $k=0$ belonging to (\textit{i}) the considered subband, (\textit{ii}) the valence subband symmetrical to it, and (\textit{iii-iv}) their respective folded continuations. Symmetrical considerations apply to the highest valence subbands. GNRs with different $N_a$ behave similarly. Since the sets of four modes described above, hereafter referred to as groups of modes, are disjoint from each other, a more efficient MS method, here called uncoupled group mode space (UGMS), can be obtained from (\ref{eqGrMStoRS})-(\ref{eqn_def_Sr_MS}) by neglecting the coupling between modes belonging to different groups. Let $\bm{v}_b(i)$ be the matrix whose columns are the modes in layer $i$ of group $b$ only. A transformation matrix $\bm{V}_b$ similar to (\ref{eqV}) is constructed by using the matrices $\bm{v}_b$ as diagonal blocks. The UGMS method is derived by further approximating (\ref{eqGrMStoRS}) as
\begin{equation}
\bm{G}^R(E) \simeq \sum_{b=1}^{N_g} \bm{V}^{}_b \bm{\widetilde{G}}^R_b(E) {\bm{V}^{}_b}^\dagger
\label{eqGrMStoRS_group}
\end{equation}
where $N_g$ is the number of the considered groups and $\bm{\widetilde{G}}^R_b$ is the solution of
\begin{equation} 
\left[ (E+\text{i}0^+) \bm{I} - \bm{\widetilde{H}}^d_b - \bm{\widetilde{\Sigma}}^R_b(E) \right]  \bm{\widetilde{G}}^R_b(E) = \bm{I}
\label{eqGrMS_group}
\end{equation} 
with
\begin{eqnarray}
\bm{\widetilde{H}}^d_{b} &=& {\bm{V}_b}^\dagger \bm{H}^d \bm{V}_b \label{eqn_def_H_MS_group} \\
\bm{\widetilde{\Sigma}}^{R}_b(E) &=& {\bm{V}_b}^\dagger \bm{\Sigma}^R(E) \bm{V}_b 
\label{eqn_def_Sr_MS_group}
\end{eqnarray}
In the UGMS method, the computational cost of the RGF algorithm scales as $O(N_g N_y^3 N_s)$, where $N_y=4$ is the number of modes in each group. Hence, if the same number of modes $N_m=4 N_g$ is used, the speed-up of UGMS compared to CMS scales as $N_g^2$, which can be sizeable for wide ribbons, since $N_g$ is proportional to the number of subbands that contribute to transport, which, in turn, is proportional to the GNR width.

In the following, we refer to the UGMS formulation, since the CMS one can be recovered from Eqs.~(\ref{eqGrMStoRS_group})-(\ref{eqn_def_Sr_MS_group}) by considering all the selected modes as belonging to the same group and by setting $N_g=1$. Note that $\bm{\Sigma}^R_S$ and $\bm{\Sigma}^R_D$ can be directly computed in MS without making use of (\ref{eqn_def_Sr_MS_group}). Instead, the calculation of $\bm{\widetilde{\Sigma}}^{(...)}_P$ is more complicated. In particular, in the MS representation these matrices are no longer diagonal, but just block diagonal.
For example, as far as $\bm{\widetilde{\Sigma}}^{<}_P$ is concerned, from the equation for $\bm{\widetilde{\Sigma}}^{<}$ analogous to (\ref{eqn_def_Sr_MS_group}), using (\ref{eqSlesserPH}) and considering for simplicity the contribution of APs only (first term at the right-hand-side), with $\bm{G}^{<}$ replaced by the equation analogous to (\ref{eqGrMStoRS_group}), the following MS expression for the element relative to modes $m$ and $m'$ of the diagonal matrix block associated with slab $i$ and group $b$ is derived
\begin{equation}
\widetilde{\Sigma}_{b, m m'}(i, i; E) = D_{ap} \sum_{b', n, n'} F^{b', n n'}_{b, m m'}(i) \, \widetilde{G}_{b', n n'} (i, i; E)
\label{eqsigmaMS}
\end{equation}
where the symbols $<$ and $P$ have been dropped for brevity and the form factor $F$ is defined by
\begin{equation}
F^{b', n n'}_{b, m m'}(i) = \sum_{\alpha} v^*_{b, \alpha m}(i) \, v^{}_{b, \alpha m'}(i) \, v^{}_{b', \alpha n}(i) \, v^*_{b', \alpha n'}(i)
\label{eqFF}
\end{equation}
($n$ and $n'$ are mode indexes within group $b'$, $\alpha$ the atom index within the slab). Identical considerations apply to the OP terms and to the other types of phonon self-energies.
It should be noticed that \textit{all the considered modes} give their contribution to $\widetilde{\Sigma}_{b, m m'}$, not only those belonging to the same group $b$ of modes $m, m'$. 
This means that intersubband scattering between all considered modes is taken into account, without the necessity of going to full RS simulations. It must also be noticed that in the solution process the different mode groups can be treated independently, which is a great advantage compared to RS and also to CMS: at every iteration pass, once the Green functions for all groups have been calculated, the self-energies are updated using the equations like (\ref{eqsigmaMS}) collecting the contributions from all modes.

In the next section the MS and RS solutions will be compared in the presence of phonon scattering. A further approximation will be tested, which consists in simplifying the form factor as
\begin{equation}
F^{b', n n'}_{b, m m'}(i) \simeq \delta_{m,m'} \delta_{n,n'} \sum_{\alpha} \left| v_{b, \alpha m}(i) \right|^2 \, \left| v_{b', \alpha n}(i) \right|^2
\label{eqFFsimple}
\end{equation}
where $\delta$ is the Kronecker delta, so that only the diagonal entries are kept in (\ref{eqsigmaMS}) and similar equations. This approximation is usually adopted in the context of the CMS EM approach due to its higher efficiency  \cite{JinAPL2006,PoliThesis2009,NikonovNH2009}.
The results obtained with expressions ($\ref{eqFF}$) and ($\ref{eqFFsimple}$) will be named ``MS2'' and ``MS1'', respectively.

\section{Results}
\label{sec_results}

All simulated devices have a double-gate structure with SiO$_2$ 1-nm-thick top and bottom oxides. As mentioned before, the source and drain are assumed to be semi-infinite leads. The portions of the source and drain regions included in the simulation domain are 10~nm long and uniformly doped. The channel is assumed to be intrinsic. Unless noted otherwise, the following device parameters are used: $N_a=13$ (corresponding to a ribbon width $W\simeq1.5$~nm), gate length $L_G=17$~nm, and doping concentration in the source and drain regions equal to $10^{-2}$ dopants per carbon atom. Also, unless stated otherwise, the MS method simulations are performed using the uncoupled group approximation (UGMS). For the reference $N_a=13$ GNR, the 2 groups of 4 modes of Fig.~\ref{fig_bandems2} are used (8 modes out of a total of 26 that correspond to the full RS solution). Both conventional FET and tunnel FET (TFET) devices are simulated. The latter have received great attention in recent years for their potential in low-power applications \cite{ZhaoNL2009}. The two structures only differ in the type of doping of the source region. While the drain is always n-type, the source is n-type and p-type for conventional FETs and TFETs, respectively.

We consider only ideal GNRs with perfect edges and no internal vacancies. Nevertheless, in the following, we also study the effect of adding a disorder potential to the TB Hamiltonian of the ideal GNR. Such disorder potential can, in principle, mimic the perturbation effect of impurities or the substrate.

\subsection{Smooth electrostatic potential}\label{sec_results_smooth}

The case of no disorder potential is treated first.

\begin{figure}
\includegraphics[width=\linewidth]{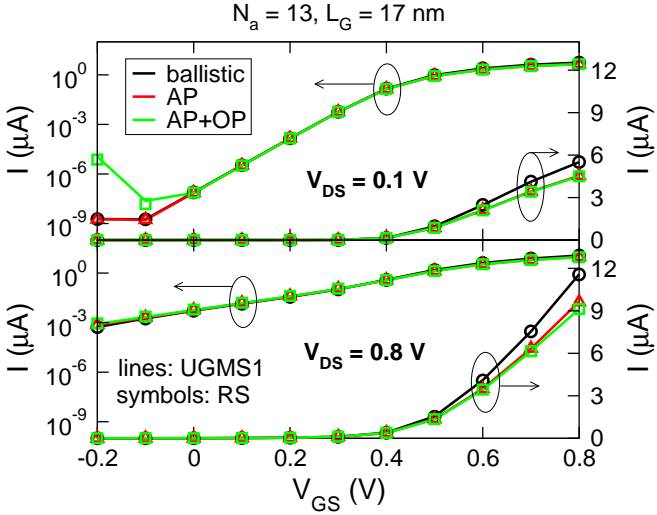}
\caption{\label{fig_turnon_nin13_L17}Turn-on characteristics of a n-i-n FET with $W=1.5$~nm at $V_{DS} = 0.1$~V (top) and $V_{DS} = 0.8$~V (bottom). Results are obtained by the RS and UGMS1 methods for different transport conditions: ballistic transport, AP scattering, and both AP and OP scattering.}
\end{figure}
The $I$ vs. $V_{GS}$ (``turn-on'') characteristics of a conventional n-i-n FET at $V_{DS}=0.1$ and 0.8~V are plotted in Fig.~\ref{fig_turnon_nin13_L17}. They are computed with the RS and the MS methods. For each method, we consider three transport conditions: (\emph{i}) without scattering (i.e. ballistic transport), (\emph{ii}) in the presence of only AP scattering, and (\emph{iii}) in the presence of both AP and OP scattering. In the case of phonon scattering and MS approach, the simplified expression of the form-factor in (\ref{eqFFsimple}) is used (MS1).

As regards the effect of phonon scattering and with reference to the RS results, it can be seen that AP scattering has only a limited effect on the current at this channel length, resulting in a ballisticity ratio (i.e. ratio between current in the presence of phonon scattering and ballistic current) of $0.8$ at $V_{GS}=0.8$~V for both $V_{DS}$ values. When also OP scattering is included, the current at high $V_{GS}$, i.e. on-state current, is only slightly decreased for the largest $V_{DS}$ value. Similar findings were reported in Ref.~\onlinecite{YoonAPL2011}. On the other hand, OP scattering is responsible for an increase of the minimum off-state current by a few orders of magnitudes when $ V_{DS}$ is low. This effect is caused by energy relaxation through emission or absorption of optical phonons, which favors band-to-band tunneling (BTBT) and shifts toward positive $V_{GS}$ values the onset of the ambipolar conduction by BTBT, similar to what occurs in carbon nanotubes \cite{KoswattaAPL2005}.

With regard to the MS results, we do not observe any significant discrepancy with respect to RS, except for the curve with OP scattering at $V_{DS}=0.1$~V, close to the $V_{GS}$ point of minimum current. This discrepancy will be discussed later in the text.

\begin{figure}
\includegraphics[width=\linewidth]{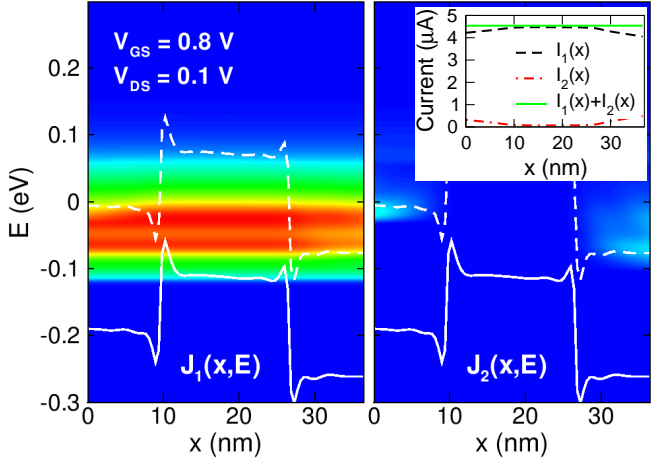}
\caption{\label{current_per_mode}Current spectrum $J_b(x,E)$ obtained by UGMS1 method for mode group $b=1$ (left) and $b=2$ (right) for the device in Fig.~\ref{fig_turnon_nin13_L17} with AP scattering and at the bias indicated in the figure. Blue (red) color means low (high) density. The white solid (dashed) line is the profile of the first (second) conduction subband. The source Fermi level is at $E=0$. The inset shows the integrals over energy $I_b(x) = \int J_b(x,E) \mathrm{d}E$ and the conservation of the total current $I_1(x)+I_2(x)$.}
\end{figure}
In order to test the validity of the MS method in describing inter-subband scattering, we separately plot in Fig.~\ref{current_per_mode} the MS current spectrum for the first and second group of modes (corresponding to the first and second subband, respectively) at $V_{GS}=0.8$~V and $V_{DS}=0.1$~V, in the case with only AP scattering. It can be seen that both subbands contribute to current. In addition, despite the absence of inelastic scattering processes in the simulation, the distribution of current over energy of the single subband is not conserved when moving from source to drain, indicating that some of the carriers are transferred from one subband to the other. Indeed, thanks to phonon scattering, part of the electrons injected from the source in the second subband reach the drain by traversing the channel in the first subband, where the energy barrier is lower. The total current is properly conserved as shown in the inset of Fig.~\ref{current_per_mode}.
\begin{figure}
\includegraphics[width=\linewidth]{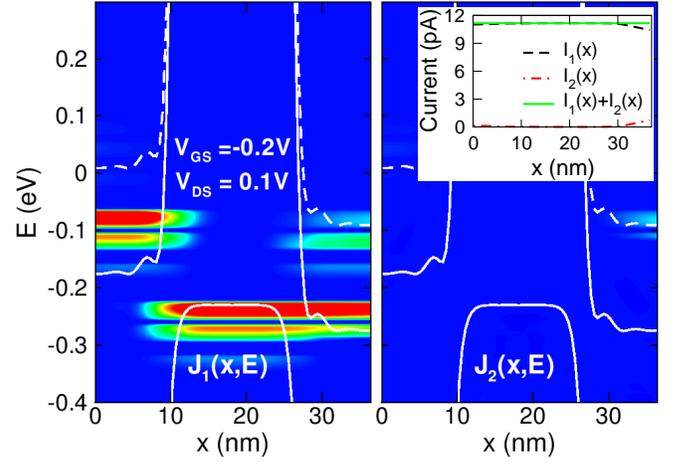}
\caption{\label{current_per_mode_AC_OPT}Same as in Fig.~\ref{current_per_mode} but with both AP and OP scattering and at $V_{GS}=-0.2$~V and $V_{DS}=0.1$~V. The white solid (dashed) lines are the profiles of the first (second) pairs of conduction and valence subbands.}
\end{figure}
Inter-subband scattering and the conservation of the total current are also evident from simulations including OP scattering: see Fig.~\ref{current_per_mode_AC_OPT}, which corresponds to a bias point where transport is dominated by phonon-assisted BTBT.

Next, we consider the TFET architecture.
\begin{figure}
\includegraphics[width=\linewidth]{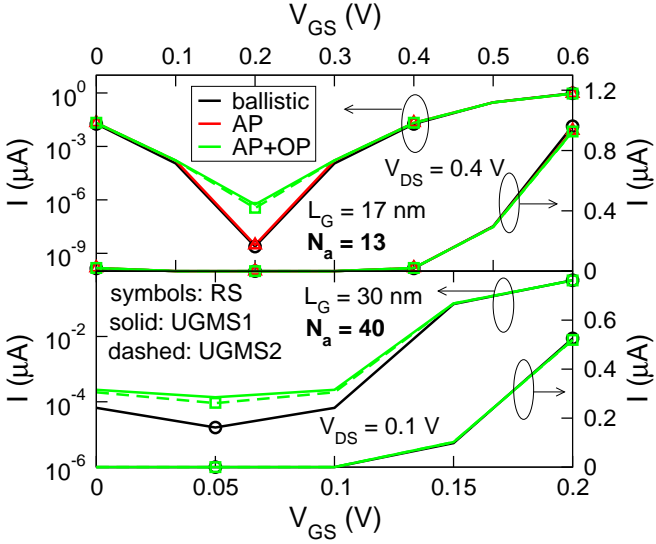}
\caption{\label{fig_turnon_pin}Turn-on characteristics of p-i-n FETs. Top: device with $W=1.5$~nm at $V_{DS} = 0.4$~V. Bottom: device with $W=5$~nm at $V_{DS} = 0.1$~V. Results are obtained by the RS and UGMS methods for different transport conditions: ballistic transport, AP scattering (only for the device with $W=1.5$~nm), and both AP and OP scattering. In the MS case, we compare the results obtained by expressions MS1 and MS2 of the form-factor.}
\end{figure}
Fig.~\ref{fig_turnon_pin}--top shows the $I$ vs. $V_{GS}$ characteristics at $V_{DS}=0.4$~V for the p-i-n counterpart of the device in Fig.~\ref{fig_turnon_nin13_L17}, calculated with the different methods and by including different types of scattering, as indicated in the legend. Results obtained with both expressions MS1 and MS2 of the form-factor are reported. The symmetry of the characteristics is related to the symmetric doping of the source and drain regions \cite{ZhaoNL2009}. By looking at the RS results, one can see that AP scattering has a negligible effect in both the on- and off-state regimes, while OP scattering significantly increases the minimum leakage current and, consequently, the minimum inverse subthreshold slope (SS), similar to the n-i-n case.
\begin{figure}
\includegraphics[width=1.0\linewidth]{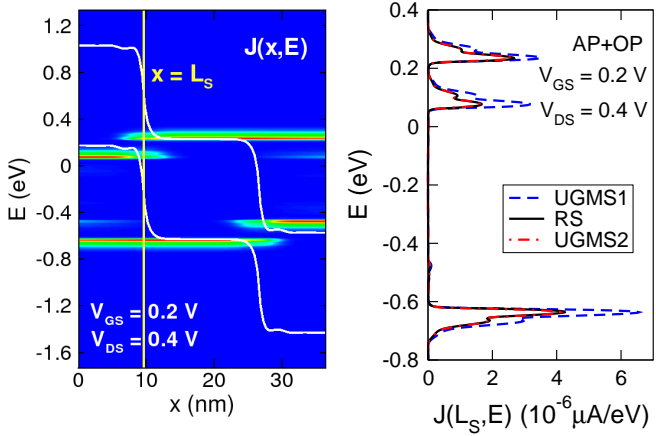}
\caption{\label{fig_jdenscut_AC_OP_Vg0_2_Vd0_4_both}Left: current spectrum $J(x,E)$ obtained by RS method for the device in Fig.~\ref{fig_turnon_pin}--top with both AP and OP scattering and at the bias indicated in the figure. The vertical line indicates the position $L_S$ of the source-channel junction. The two white lines are the profile of the first conduction and valence subbands. Right: comparison between the current spectra at $x=L_S$ obtained by RS, UGMS1, and UGMS2 methods.}
\end{figure}
As illustrated by the current spectrum in Fig.~\ref{fig_jdenscut_AC_OP_Vg0_2_Vd0_4_both}--left, the increase of the minimum current is due to phonon-assisted BTBT at the source-channel and drain-channel junctions: although the first conduction and valence subbands do not face each other, electrons can transmit from the valence to the conduction subband by absorption of optical phonons.
These results are in agreement with the ones in Ref.~\onlinecite{YoonAPL2012}.

Again, the only difference (of about a factor of $1.5$) between the RS and MS1 results is noticed in the presence of OP scattering at the point of minimum current. However, the accuracy with respect to RS can be almost completely recovered by the MS2 method, which uses the exact expression of the form-factor in (\ref{eqFF}). This is more clearly shown by the comparison in Fig.~\ref{fig_jdenscut_AC_OP_Vg0_2_Vd0_4_both}--right between the current spectra at the source-channel junction obtained with the different methods. The lack of accuracy of (\ref{eqFFsimple}) in this bias condition could be related to the neglect of the terms $F^{b, m n}_{b, m n}(i)$ with $n \neq m$, which are actually of the same size as the terms $F^{b, n n}_{b, m m}(i)$ included in (\ref{eqFFsimple}). As a drawback, a slow-down of the simulation by about a factor of $1.5$ has been measured using MS2 compared to MS1, which can be ascribed to the increased computational cost of (\ref{eqsigmaMS}).

The above considerations apply also to the turn-on characteristics of a wider GNR TFET in Fig.~\ref{fig_turnon_pin}--bottom. For this device, a lower $V_{DS}$ is chosen due to the lower band gap. In addition, we set $N_a=40$ (corresponding to $W\simeq5$~nm), $L_G=30$~nm, and a doping concentration in the source and drain regions of $7 \cdot 10^{-4}$ dopants/atom. 4 groups of 4 modes are used in the UGMS simulations. It is worth noticing that the UGMS method is still accurate for the wider ribbon, indicating that the decoupling in separate groups is still valid, even though the subbands are more closely spaced than in the $N_a=13$ GNR.

\begin{figure}
\includegraphics[width=\linewidth]{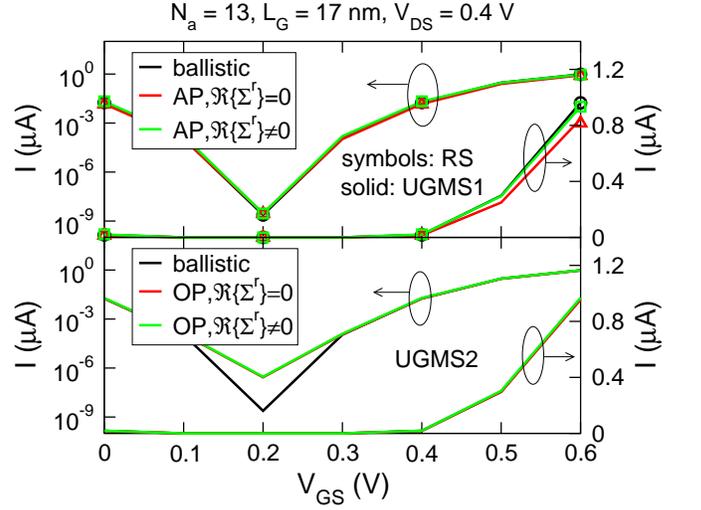}
\caption{\label{fig_turnon_pin13_Vd0_4_realpart}Turn-on characteristics for the same device as in Fig.~\ref{fig_turnon_pin}--top comparing different approximations for $\bm{\Sigma}^R_P$. Top: only AP scattering is included, with or without the Hermitian part of $\bm{\Sigma}^R_P$. Bottom: only OP scattering is included, with or without the Hermitian part of $\bm{\Sigma}^R_P$. The ballistic curve is shown for reference in both figures and the solution method for each curve is indicated in the legend.}
\end{figure}

To evaluate the importance of the Hermitian part of $\bm{\Sigma}^R_P$, here denoted by $\Re\{ \bm{\Sigma}^R_P \}$, we report in Fig.~\ref{fig_turnon_pin13_Vd0_4_realpart}
the $I$ vs. $V_{GS}$ characteristics of the $N_a=13$ TFET computed with or without $\Re\{ \bm{\Sigma}^R_P \}$, separately for each type of scattering.

\begin{figure}
\includegraphics[width=\linewidth]{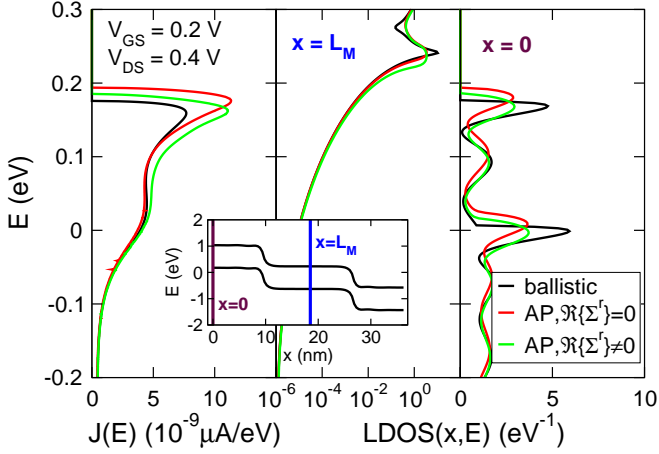}
\caption{\label{fig_jdens_AC_Vg0_2_Vd0_4}Current spectrum (left) and integral over the slab of the LDOS at $x=L_M$ (center) and $x=0$ (right) for the device in Fig.~\ref{fig_turnon_pin13_Vd0_4_realpart}--top at $V_{GS} = 0.2$~V and $V_{DS} = 0.4$~V. $L_M$ is the mid-channel position. RS method is used. The ballistic subband profile is shown in the inset.}
\end{figure}

For AP scattering (Fig.~\ref{fig_turnon_pin13_Vd0_4_realpart}--top), the neglect of $\Re\{ \bm{\Sigma}^R_P \}$ leads to an underestimation of the on-state current. This can be understood as follows. In general, $\Re\{ \bm{\Sigma}^R_P \}$ has the effect of shifting the Hamiltonian eigenvalues \cite{DattaETMS1997}. However, in GNRs, the shift is of opposite sign for energies above and below the GNR mid-gap due to the symmetry of the subband structure. The result is a decrease of the GNR band gap, which favors BTBT, so that a larger current is expected when $\Re\{ \bm{\Sigma}^R_P \}$ is included in the simulation. Interestingly, the two expressions of $\Re\{\bm{\Sigma}^R_P\}$ give almost identical values with respect to the minimum current. To clarify this point, we plot in Fig.~\ref{fig_jdens_AC_Vg0_2_Vd0_4}
the current spectrum and the local density of states (LDOS) per slab at two positions along the device, for the bias point corresponding to the minimum current. First, it can be seen that the current spectrum with AP scattering and $\Re\{ \bm{\Sigma}^R_P \} \neq 0$ is larger than the ballistic one: the reason can be ascribed to an enhanced BTBT through the channel region due to the band gap narrowing effect mentioned above, which can be appreciated from the logarithmic plot of the LDOS at the mid-channel position in Fig.~\ref{fig_jdens_AC_Vg0_2_Vd0_4}--center. Secondly, by looking at Fig.~\ref{fig_jdens_AC_Vg0_2_Vd0_4}--right, it can be noticed that the peaks of the LDOS in the source region with $\Re\{ \bm{\Sigma}^R_P \} \neq 0$ are located at the same energy positions as the ones of the ballistic LDOS. On the contrary, the peaks of the LDOS with $\Re\{ \bm{\Sigma}^R_P \} = 0$ are shifted up in energy, resulting in a tunneling current larger than the ballistic one and similar to the one with $\Re\{ \bm{\Sigma}^R_P \} \neq 0$ (Fig.~\ref{fig_jdens_AC_Vg0_2_Vd0_4}--left). The shift of the LDOS can be attributed to a ``loss of charge'' when $\Re\{ \bm{\Sigma}^R_P \}$ is set to zero \cite{SvizhenkoPRB2005} and to the combined effect of the electrostatic feedback.

For OP scattering instead, no relevant difference is observed when including $\Re\{ \bm{\Sigma}^R_P \}$, even at high $V_{GS}$ (Fig.~\ref{fig_turnon_pin13_Vd0_4_realpart}--bottom).

\subsection{Disordered electrostatic potential}\label{sec_results_disorder}

We focus on the $N_a=13$ TFET. The simulations are performed in a non-self-consistent way, by solving the NEGF equations with a fixed electrostatic potential. We take the electrostatic potential as the sum of a disorder potential and the one calculated self-consistently in the absence of disorder and in the ballistic limit at $V_{GS}=0.6$~V and $V_{DS}=0.4$~V, using the RS approach. Two types of disorder are considered: a ``long-range'' one, i.e. slowly varying on the atomic scale, and a ``short-range'' one, i.e. rapidly varying from one atom to its neighbor ones. According to the first model, which is derived from Ref.~\onlinecite{PoljakTED2012}, the disorder energy potential $V_i$ at the atomic site $i$, located at position $\vec{r}_i$, is calculated as
\begin{equation}
V_i = \sum_{j=1}^N S_j \delta V \exp \left( - \frac{\left| \vec{r}_i - \vec{X}_j \right|^2}{l^2} \right)
\end{equation}
where $N$, $l$, $\delta V$ are parameters, while $S_j$ and $\vec{X}_j$ are random variables: $S_j$ can take values $\pm 1$ with equal probability; $\vec{X}_j$ is uniformly distributed over all the atomic position $\vec{r}_i$. The second model we study is the Anderson type of disorder \cite{LherbierPRL2008}, according to which $V_i=Y_i$, where $Y_i$ is a random variable uniformly distributed in $[-\delta V /2, \delta V /2]$. 

The coherent case, i.e. without phonon scattering, is considered first. 
\begin{figure}
\includegraphics[width=\linewidth]{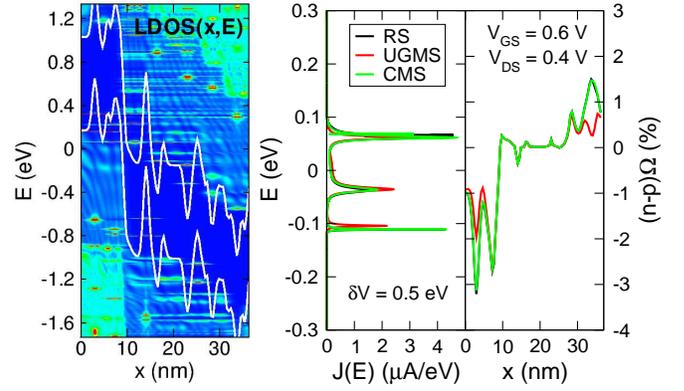}
\caption{\label{fig_ldos_jdens_rho_DIS_BAL_new}Integral over the slab of the LDOS (left), current spectrum (center), and average over the slab of the net electron concentration per carbon atom (right) for the device in Fig.~\ref{fig_turnon_pin}--top at $V_{GS} = 0.6$~V and $V_{DS} = 0.4$~V, in the presence of a long-range disorder potential and in the coherent limit. LDOS is obtained by RS, whereas the solution method for the other curves is indicated in the legend. $\Omega=3 a_{cc}^2/4$ is half the area of the graphene unit cell.}
\end{figure}
In Fig.~\ref{fig_ldos_jdens_rho_DIS_BAL_new}--left we show the LDOS corresponding to a realization of the long-range type of disorder, obtained with $N=0.01 N_c$, $l=5 a_{cc}$, and $\delta V=0.5$~eV, where $N_c$ is the total number of carbon atoms inside the device and $a_{cc}$ is the carbon-carbon bond length. The resonant states induced by disorder are clearly visible in the LDOS. In Fig.~\ref{fig_ldos_jdens_rho_DIS_BAL_new}--center and Fig.~\ref{fig_ldos_jdens_rho_DIS_BAL_new}--right we compare the current spectrum and net electron charge along the device, respectively, obtained with RS, UGMS, and CMS. The UGMS simulations are performed with the same 2 groups of 4 modes used in the case without disorder, while the CMS ones use the same 8 modes but all coupled in one group. It turns out that the UGSM method looses accuracy due to disorder-induced mode mixing, while the CMS method leads to results very close to the RS ones (e.g. the error on the current value is less than $2\%$).

\begin{figure}
\includegraphics[width=\linewidth]{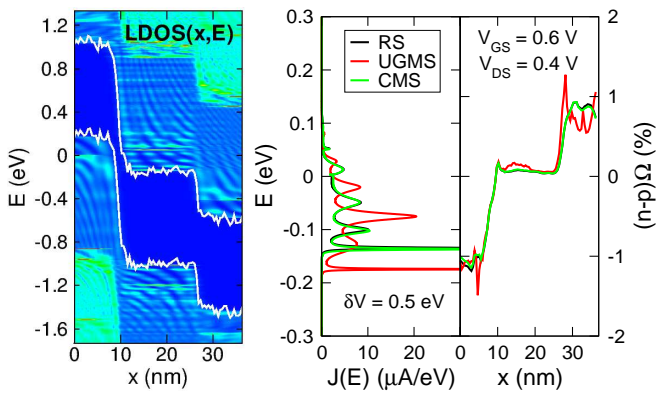}
\caption{\label{fig_ldos_jdens_rho_AND_BAL_new}Same as in Fig.~\ref{fig_ldos_jdens_rho_DIS_BAL_new} but in the presence of a short-range disorder potential.}
\end{figure}
Similar considerations can be made regarding the results obtained with the second model of disorder with $\delta V = 0.5$~eV. (Fig.~\ref{fig_ldos_jdens_rho_AND_BAL_new}). The more regular LDOS pattern and the higher current spectrum in Fig.~\ref{fig_ldos_jdens_rho_AND_BAL_new} compared to Fig.~\ref{fig_ldos_jdens_rho_DIS_BAL_new} indicate that the amount of scattering in this device is smaller.

\begin{figure}
\includegraphics[width=\linewidth]{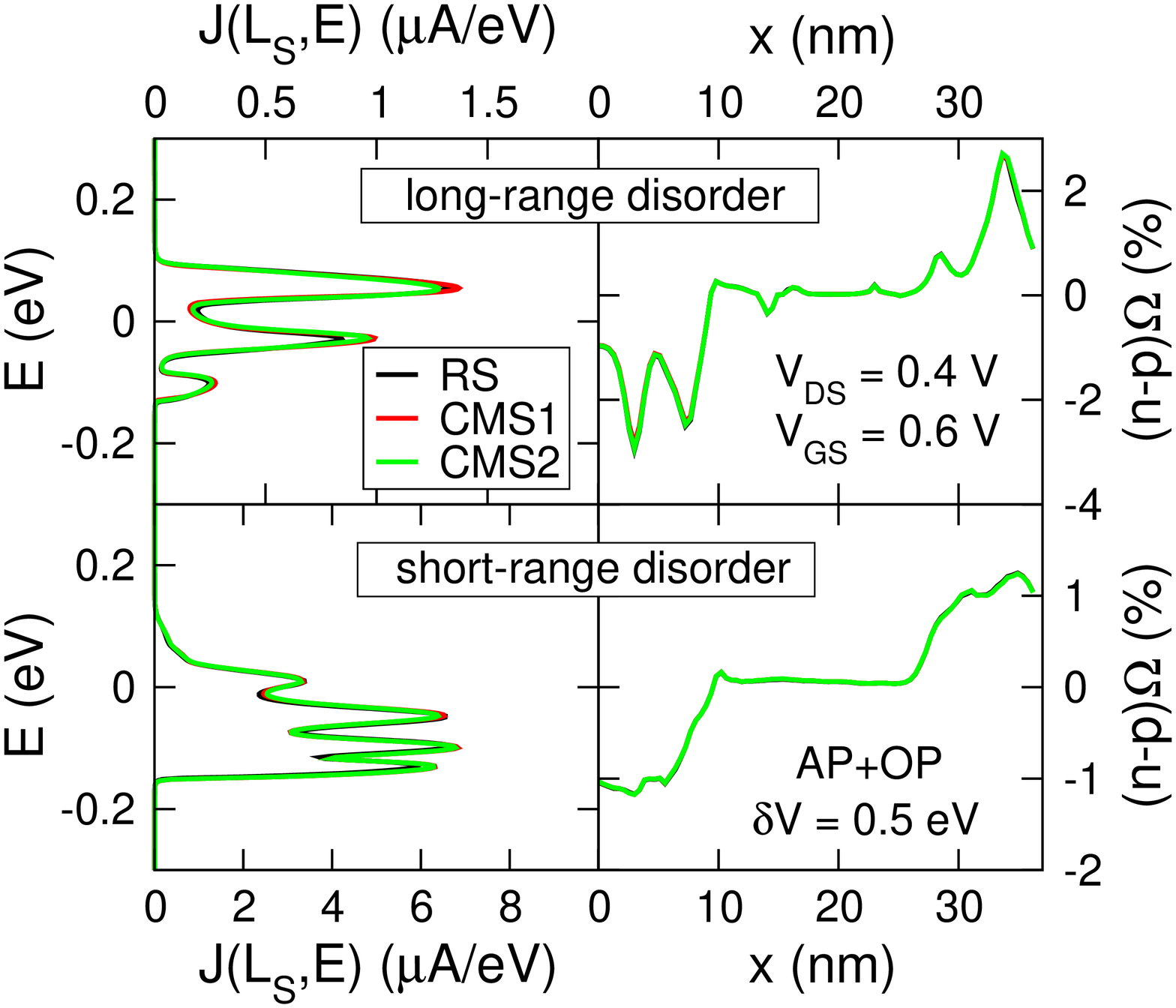}
\caption{\label{fig_jdens_rho_DIS_AC_OPT}Top: current spectrum at the source-channel junction (left) and average over the slab of the  net electron concentration per carbon atom (right) for the device in Fig.~\ref{fig_turnon_pin}--top at $V_{GS} = 0.6$~V and $V_{DS} = 0.4$~V, in the presence of a long-range disorder potential and with both AP and OP scattering. The solution method for each curve is indicated in the legend. Bottom: same as in (top) but in the presence of a short-range disorder potential.}
\end{figure}
The simulations have been repeated including AP and OP scattering (Fig.~\ref{fig_jdens_rho_DIS_AC_OPT}). For both types of disorder, it is seen that the MS1 and MS2 approximations of the form-factor provide similar results and that the CMS approach is still accurate with respect to RS. By comparing the current spectra in Fig.~\ref{fig_jdens_rho_DIS_AC_OPT} with the ones in Figs.~\ref{fig_ldos_jdens_rho_DIS_BAL_new}--\ref{fig_ldos_jdens_rho_AND_BAL_new}, it can be noticed that phonon scattering adds a significant broadening. In the case of short-range disorder, phonon scattering only slightly decreases the current (from $0.83$ to $0.80$~$\mu$A). On the other hand, in the case of long-range disorder, the current is increased from $0.084$ to $0.10$~$\mu$A when phonon scattering is included in the simulation, indicating that the localization transport regime \cite{DattaETMS1997} that occurs in the coherent approximation is broken by the dephasing effect of phonon scattering.

\section{Conclusions}
\label{sec_conclusions}

A mode space method for TB NEGF simulations of armchair GNR FETs including phonon scattering has been presented and tested with reference to both conventional and tunnel FET structures. When no disorder is included in the simulation, an efficient decoupling of the modes in different groups (UGMS) can be employed with excellent accuracy. Despite the decoupling, the method correctly accounts for inter-subband scattering. Simplified expressions of the scattering-self energies have been compared. The one obtained by neglecting some entries of the form-factor is found to be accurate except for the bias points where transport occurs by phonon-assisted BTBT. On the other hand, the real part of the scattering self-energy has only a limited effect on the device characteristics, especially for the case of optical phonon scattering, where its calculation is most demanding. In the presence of a disorder potential, the modes need to be coupled in a single group (CMS) to account for mode mixing, but no additional modes, compared to the ones used to simulate the case without disorder, need to be included to achieve accurate results.

While the computational advantage of CMS over RS is about a constant factor with respect to the GNR width (equal to about 30 for the simulation parameters chosen in this paper), the speed-up of UGMS compared to RS is about $40$ for a device width of 1.5~nm and increases proportionally to the second power of the GNR width (for the 5-nm-wide device considered in this paper such speed-up is about $360$).

\begin{acknowledgments}
R.~G. would like to thank Dr.~E.~Baravelli of University of Bologna for fruitful discussions on the mode-space approach. This work has been supported by the EU project GRADE 317839. The authors acknowledge the CINECA Award N. HP10CPFJ69, 2011 for the availability of high performance computing resources and support.
\end{acknowledgments}

\appendix

\section{Calculation of the retarded phonon self-energy} \label{sec_hermitian_part}

We consider here only the RS case, the generalization of the expressions to MS being straightforward.

Replacing (\ref{eqSlesserPH}) and (\ref{eqSgreaterPH}) into the last member of (\ref{eqSrPH}) and using the identity analogous to (\ref{eqSrPH}) valid for the $\bm{G}$ matrices, one can write
\begin{equation} 
\bm{\Sigma}^R_P(E) = \bm{\Sigma}^R_{AP}(E)+\bm{\Sigma}^R_{OP}(E)
\label{eqSrPH2}
\end{equation}
with 
\begin{equation}
\bm{\Sigma}^R_{AP}(E) =  D_{ap} \bm{I} \circ \bm{G}^R(E)
\label{eqSrAP}
\end{equation}
and
\begin{eqnarray}
&&\bm{\Sigma}^R_{OP}(E) = D_{op} \bm{I} \circ \bigg\{ \nonumber\\
&& (N_{op}+1) \bm{G}^R(E-\hbar \omega_{op}) +
 N_{op} \bm{G}^R(E+\hbar \omega_{op}) + \nonumber\\
&&+\frac{\bm{G}^<(E-\hbar \omega_{op}) - \bm{G}^<(E+\hbar \omega_{op} )}{2} + \nonumber\\
&&+ \text{i} \text{P} 
\! \! \! \int\limits_{-\infty}^{+\infty} {{\bm{G}^<(E^\prime -\hbar \omega_{op}) - \bm{G}^<(E^\prime+\hbar \omega_{op})} 
\over{2\pi(E - E^\prime})} \text{d}E^\prime \bigg\} 
\label{eqSrOP}
\end{eqnarray}
which does not contain $\bm{G}^>$. It is worth noticing that the principal part integral in (\ref{eqSrOP}) contains only a fraction of the Hermitian part of $\bm{\Sigma}^R_{OP}$ \citep{EspositoJCE2009}. It is calculated here by means of a piecewise constant approximation of $\bm{G}^<(E^\prime)$ over the energy domain $[E_{\text{min}},E_{\text{max}}]$, which is discretized uniformly with energy steps $\Delta_E$ typically of the order of $\hbar \omega_{op}/100$, i.e.
\begin{eqnarray} 
&\text{P}& \! \! \! \int\limits_{-\infty}^{+\infty} {{\bm{G}^<(E^\prime \mp \hbar \omega_{op})} 
\over{E - E^\prime}} \text{d}E^\prime \simeq\nonumber\\
&\simeq& \sum_j \bm{G}^<(E_j) \ln 
\left|{{E-E_j+\Delta_E/2 \mp \hbar \omega_{op}}\over{E-E_j-\Delta_E/2 \mp \hbar \omega_{op}}}\right|
\label{eqIntDiscr}
\end{eqnarray}
where the summation extends over all discrete energy points $E_j \in [E_{\text{min}},E_{\text{max}}]$. An expression of $\bm{\Sigma}^R_{OP}$ analogous to (\ref{eqSrOP}) containing $\bm{G}^>$ instead of  $\bm{G}^<$ can also be derived. Expression (\ref{eqSrOP}) is preferred for $E$ higher than the contact Fermi energies, since the numerical error introduced by truncating the upper limit of the integral to $E_{\text{max}}$ is minimized due to the decaying nature of $\bm{G}^<$ at high energies. On the contrary, the alternative expression of $\bm{\Sigma}^R_{OP}$ which depends on $\bm{G}^>$ is used for low energies below the contact Fermi levels, for analogous reasons. For intermediate energies an average of the two formulations is used. 

In Sec.~\ref{sec_results_smooth}, the results obtained with the full expression of $\bm{\Sigma}^R_{AP}$ and $\bm{\Sigma}^R_{OP}$ are compared with approximate solutions obtained by neglecting the respective Hermitian parts, i.e.
\begin{equation} 
\bm{\Sigma}^R_{AP}(E) \simeq D_{ap} \bm{I} \circ {{\bm{G}^>(E) - \bm{G}^<(E)} \over {2}} 
\label{eqSrAPsimpl}
\end{equation}
and
\begin{eqnarray} 
&&\bm{\Sigma}^R_{OP}(E) \simeq \frac{D_{op}}{2} \bm{I} \circ \bigg\{ (N_{op}+1) \bm{G}^>(E-\hbar \omega_{op}) + \nonumber\\
&&+ N_{op} \bm{G}^>(E+\hbar \omega_{op}) - (N_{op}+1) \bm{G}^<(E+\hbar \omega_{op}) + \nonumber\\
&&- N_{op} \bm{G}^<(E-\hbar \omega_{op}) \bigg\} 
\label{eqSrOPsimpl}
\end{eqnarray}
Expression (\ref{eqSrOPsimpl}) is much more efficient than (\ref{eqSrOP}) due to the absence of the integral term.  


%
%

%


\bibliography{mybibfile}

\end{document}